\documentclass[prb,twocolumn,floatfix]{revtex4}
\usepackage{graphicx}
\usepackage{color}
\usepackage{amsmath}
\usepackage{amssymb}
\usepackage{latexsym}
\usepackage{psfrag}

\begin{document}

\title{Spin polarization in modulation-doped GaAs quantum wires}
\author{M. Evaldsson, S. Ihnatsenka and I. V. Zozoulenko}
\affiliation{Solid State Electronics, Department of Science and Technology (ITN), Link\"{o}ping University, 601 74 Norrk\"{o}ping, Sweden} \date{\today}

\begin{abstract}
We study spin polarization in a split-gate quantum wire focussing on the effect of a realistic smooth potential due to remote donors.  Electron interaction and spin effects are included within the density functional theory in the local spin density approximation. We find that depending on the electron density, the spin polarization exhibits qualitatively different features. For the case of relatively high electron density, when the Fermi energy $E_{F}$ exceeds a characteristic strength of a long-range impurity potential $V_{donors}$, the density spin polarization inside the wire is practically negligible and the wire conductance is spin-degenerate. When the density is decreased such that $E_{F}$ approaches $V_{donors}$, the electron density and conductance quickly become spin polarized. With further decrease of the density the electrons are trapped inside the lakes (droplets) formed by the impurity potential and the wire conductance approaches the pinch-off regime. We discuss the limitations of DFT-LSDA in this regime and compare the obtained results with available experimental data.
\end{abstract}

\maketitle

%*********************************************
%*************** ABSTRACT ********************
%*********************************************

%*************************************************
%***************** INTRODUCTION ******************
%*************************************************

\subsection*{Introduction}

The possibility of a spontaneous spin-polarization at low electron densities in low dimensional electron systems has attracted an enormous interest over the past years. The phenomena has been suggested to occur in various systems including quantum point contacts \cite{Thomas1996} (QPCs), two-dimensional electron gas (2DEG) \cite{Ghosh2004, Goni2004}, quantum wire \cite{Auslaender2005} and open quantum dots \cite{QD_spin}. Theoretical modeling with Hartree-Fock\cite{Andreev1998}, the density functional theory \cite{Wang, Ferry2005, Ihnatsenka2007,QD_spin} and Monte Carlo simulations\cite{Varsano2001} has reproduced low density spin polarization in a number of systems. The mechanism driving the polarization is the exchange energy dominating over the kinetic energy at low densities, making the spin-polarized state the energetically most favorable. The electron density necessary for this polarization to occur is generally very low, below $ n_{s}\sim 3\times 10^{14}$m$^{-2}$ (corresponding to the interaction parameter\cite{Reimann_RevModPhys} $r_{s}=1/a_{B}^{\ast }\sqrt{\pi n_{s}}\approx 3.2$, where $a_{B}^{\ast }$ is the effective Bohr radius) as indicated in, e.g., \cite{Goni2007, Ghosh2007}. At such low electron densities the electrostatic potential due to impurities in the donor layer can significantly affect the electronic and transport properties of the 2DEG. For example, Nixon \emph{et al.}\cite{Davies1990} showed that a monomode quantum wire is difficult to achieve because of the pinch-off due to a random potential from unscreened donors. This pinch-off is characterized by the critical electron density $n_{c}$ , -- the density where the 2DEG undergoes a metallic-insulator transition (MIT)\cite{Davies1990, Efros1993, Ilani2000, Baenninger2005}. The MIT causes a localization of the electron gas with an accompanying abrupt change in the conductance\cite{Ilani2000, Baenninger2005}. Recent measurements of the thermodynamic magnetization in silicon 2DEG:s\cite{Prus,Shashkin} found an enhancement of the spin susceptibility close to $n_{c}$. Interestingly, theoretical considerations \cite{Andreev1998, Punnoose} indicate that the spin-susceptibility in a disordered potential \emph{increases} for electron densities close to $n_{c}$. These studies considered the general behavior of a 2DEG in an impurity potential but did not elaborate on specific geometries, e.g., quantum dots or wires.

In the present paper we study spin polarization in a split-gate quantum wire focussing on the effect of a realistic smooth potential due to remote donors. For this purpose we, starting from a heterostructure and a gate layout, model GaAs/AlGaAs quantum wires within the density functional theory in the local spin density approximation (DFT-LSDA) accounting for a long-range impurity potential due to ionized dopants. A gate voltage applied to a top gate allow us to tune the electron density in the wire close to $n_{c}$. We find that depending on the electron density, the spin polarization exhibits qualitatively different features. For the case of relatively high electron density, when the Fermi energy $E_{F}$ exceeds a characteristic strength of a long-range impurity potential $V_{donors}$, the density spin polarization inside the wire is practically negligible and the wire conductance is spin-degenerate. When the density is decreased such that $E_{F}$ approaches $V_{donors},$ the electron density and conductance quickly become spin polarized. With further decrease of the density the electrons are trapped inside the lakes (droplets) formed by the impurity potential and the wire conductance approaches the pinch-off regime.

%
%*************************************************
%******************** MODEL **********************
%*************************************************
%

\subsection*{Model}

We study the conductance and electron density of a GaAs/AlGaAs quantum wire with a realistic long-range impurity potential due to remote donors.
%*************************************************
%************* DESCRIPTION OF SYSTEM *************
%*************************************************
%
%*************************************************
%****************** FIGURE ONE *******************
%*************************************************
\begin{figure}[tbh]
\includegraphics{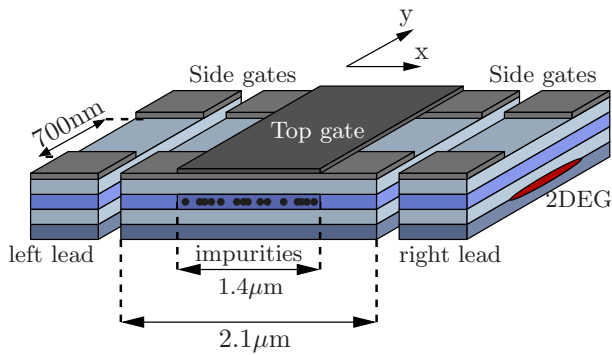}
\caption{(color online) Schematic view of the system studied. The heterostructure consists of (from bottom to top), a GaAs substrate, a 30-60nm AlGaAs-spacer, a 26nm donor layer and a 14nm cap layer. The side gates define a quantum wire and the top gate controls the electron density in the section with randomly distributed dopants $\protect\rho _{d}(\mathbf{r})$.}
\label{fig:figure1}
\end{figure}
In the wire, sketched in Fig.\ 1, the confinement is induced by two metallic side gates situated 700 nm apart on the top of the heterostructure. The heterostructure consists of the cap layer, the donor layer and the spacer. Electrons from the fully ionized donor layer form a two-dimensional electron gas at a GaAs/AlGaAs interface situated at the distance $d_{2DEG}$ below the surface. The confining potential from the donor layer is different in the leads and in the central region of the device. The leads, which extend to electron reservoirs at infinity, are considered ideal and the ionized donors in the lead regions are treated as a uniform layer with the density $\rho _{d}$ at the distance $d_{d}$ from the surface. Thus, the leads guides charges from the reservoirs to and from the middle region without any scattering. In the middle section the \emph{average} donor density is still $\rho _{d}$, but a gradual transition from the uniform donor density in the leads to random placement of the dopants is implemented. As a results, the electrons are scattered in the middle region of the wire due to the long-range impurity potential. The donor potential on the depth of the 2DEG is calculated according to the procedure outlined by Davies \textit{et al.} \cite{Davies1990}. An additional top gate in the middle section allows us to control the electron density in this region. The potential on the top gate, $V_{g}$, ranges from zero voltage up to $-0.07$ V, which is the pinch off voltage for at least one of the spin species. The self-consistent potential at the GaAs/AlGaAs interface for different gate voltages is illustrated in Fig.\ \ref{fig:figure2}. %
%*************************************************
%***************** FIGURE TWO *******************
%*************************************************
\begin{figure}[tbh]
\includegraphics{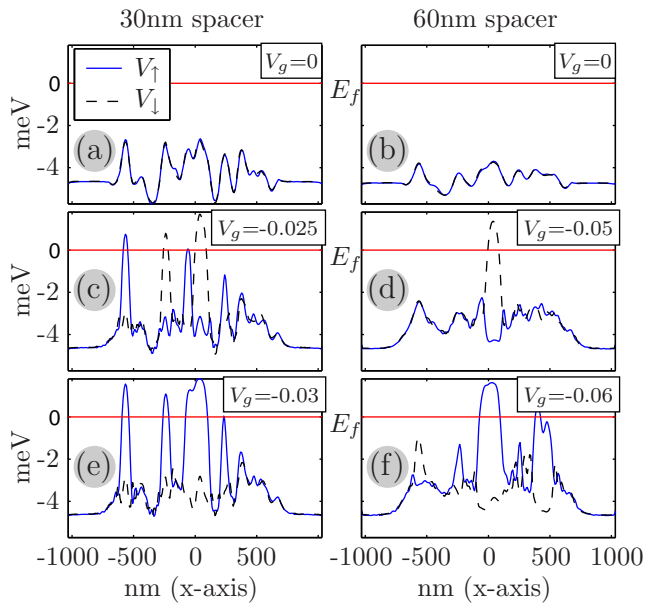}
\caption{(color online) The self consistent spin-up/down potential along a slice in the middle of the wire ($y=0$) for increasing gate voltages, $V_{g}$ . (a), (c) and (e) show the 30nm-spacer sample while (b), (d) and (f) the 60nm-spacer.}
\label{fig:figure2}
\end{figure}
%*************************************************
%****************** HAMILTONIAN ******************
%*************************************************
The shape of the impurity potential depends on the spacer thickness. We consider two cases when the width of the spacer layer is respectively 30 nm and 60 nm. In both cases the \emph{average} electron density is similar ($n^{\uparrow }+n^{\downarrow }\sim 1\times 10^{15}$m$^{-2}$ at $V_{g}=0$). This is achieved by choosing slightly different impurity concentrations for different spacer layers ($\rho _{imp}=1\times 10^{24}$m$^{-3}$ and $\rho _{imp}=1.07\times 10^{24}$m$^{-3}$ for respectively 30nm and 60nm spacers). The effective width of the wire in the 2DEG is 500 nm for both cases. For the case of 60nm spacer, the donors in the central region are situated further away from the 2DEG which results in a smoother profile in comparison to the 30nm case, c.f. Fig.\ \ref{fig:figure2} (a),(c),(d) and (b),(d),(f).

Using the Kohn-Sham formalism we write the Hamiltonian for the quantum wire as \cite{Vignale}
\begin{equation}
H^{\sigma }=-\frac{\hbar ^{2}}{2m^{\ast }}\nabla ^{2}+V^{\sigma }(\mathbf{r})
\label{eq:hamiltonian}
\end{equation}%
where $m^{\ast }=0.067m_{e}$ is the effective mass in GaAs, $\mathbf{r}=(x,y) $ and $\sigma $ stands for spin up/down electrons ($\uparrow ,\downarrow $). The total potential $V^{\sigma }(\mathbf{r})$ can be written as the sum of the classical Hartree potential, $V_{H}(\mathbf{r})$, the correlation and exchange potential, $V_{xc}^{\sigma }(\mathbf{r})$, and the external potential due gates, donors and Schottky barrier, $V_{ext}(\mathbf{r})$. %*************************************************
%****************** POTENTIALS *******************
%*************************************************
\begin{equation}
V^{\sigma }(\mathbf{r})=V_{H}(\mathbf{r})+V_{xc}^{\sigma }(\mathbf{r} )+V_{ext}(\mathbf{r})  \label{eq:potential}
\end{equation}
With mirror charges at distance $d_{2DEG}$ above the surface the Hartree potential is written as
\begin{multline}
V_{H}(\mathbf{r})=\frac{e^{2}}{4\pi \epsilon \epsilon _{0}}\int d\mathbf{r} ^{\prime }n(\mathbf{r}^{\prime })\Bigg(\frac{1}{|\mathbf{r}-\mathbf{r^{\prime }}|}  \label{eq:hartree_potential} \\ -\frac{1}{\sqrt{|\mathbf{r}-\mathbf{r^{\prime }}|^{2}+4d_{2DEG}^{2}}}\Bigg)
\end{multline}%
where $n(\mathbf{r})$ is the total ($n^{\uparrow}+n^{\downarrow}$)
electron density. Within the LSDA approximation the exchange and correlation potential is given by
\begin{equation}
V_{xc}^{\sigma }(\mathbf{r})=\frac{\delta }{\delta n^{\sigma}}(n\varepsilon_{xc}(n)).  \label{eq:xc_potential}
\end{equation}
For $\varepsilon _{xc}$ the parametrization by Tanatar and Ceperly\cite{TC} was implemented. Finally, for $V_{ext}(\mathbf{r})=V_{gates}(\mathbf{r})+ V_{donors}(\mathbf{r})+V_{Schottky}(\mathbf{r})$ we use analytical expressions for $V_{gates}(\mathbf{r})$\cite{Davies1995} and $V_{donors}( \mathbf{r})$ (Refs. \cite{Martorell} and \cite{Davies1990} respectively for the lead- and middle sections of the wire); the Schottky barrier $V_{Schottky}(\mathbf{r})$ is set to 0.8eV.
%*************************************************
%********* GREEN'S FUNCTION DEFINITION ***********
%*************************************************
Using the recursive Green's function technique with mixed basis set\cite{Zozoulenko} we compute the conductance through the scattering region (middle section) and the self-consistent electron density in the system. Details of our implementation can be found in \cite{Ihnatsenka2005,Ihnatsenka2007,open_dot} and the procedure will only be briefly sketched here. The Hamiltonian Eq.~(\ref{eq:hamiltonian}) is discretized on an equidistant grid and the retarded Green's function is defined as
\begin{equation}
\mathcal{G}^{\sigma }=(E-H^{\sigma }-i\epsilon )^{-1}.
\label{eq:greensfunction}
\end{equation}%
The electron density is integrated from the Green's function (in the real space),
\begin{equation}
n^{\sigma }=-\frac{1}{\pi }\int_{-\infty }^{\infty }\Im \lbrack \mathcal{G}^{\sigma }(\mathbf{r},\mathbf{r},E)]f_{FD}(E-E_{F})dE,
\label{eq:electrondensity}
\end{equation}%
%
%
%
%
%
%
%
%
%
%
%*************************************************
%************ COMPUTATIONAL PROCEDURE ************
%*************************************************
$f_{FD}$ being the Fermi-Dirac distribution. First we compute the self-consistent solution of equations (\ref{eq:hamiltonian})-(\ref{eq:electrondensity}) for an infinite homogenous wire by the technique described in \cite{Ihnatsenka2005}. The converged solution for the infinite wire is used to find an approximation for the surface Green's function of the left and right leads. This approximation can be justified because of a sufficient separation between the leads and the inhomogeneous potential in the middle section such that any inhomogeneous contribution to the potential from the middle section is negligible at the leads. Next we apply the Dyson equation to couple the left and right surface Green's function and recursively compute the full Green's function for the middle section. We then iterate Eq.~(\ref{eq:hamiltonian})-(\ref{eq:electrondensity}) to find a self-consistent solution for the middle section. On each iteration step $i$ the electron density is updated from the input and output densities of the previous step, $n_{i+1}^{in}(\mathbf{r})=(1-\varepsilon )n_{i}^{in}(\mathbf{r})+ \varepsilon n_{i+1}^{out}(\mathbf{r}),$ $\varepsilon $ being a small number, $\sim 0.05$. Convergence is defined as a ratio between the relative change in input/output density at the iteration step $i$, $\left\vert n_{i}^{out}-n_{i}^{in}\right\vert /(n_{i}^{out}+n_{i}^{in})<10^{-5}$ .
%*************************************************
%****************** CONDUCTANCE ******************
%*************************************************
Finally the conductance is computed from the Landauer formula, which in the zero bias limit is
\begin{equation}
G^{\sigma }=-\frac{e^{2}}{h}\int_{-\infty }^{\infty }dET^{\sigma }(E)\frac{\partial f_{FD}(E-E_{F})}{\partial E}  \label{eq:landauer}
\end{equation}%
where $T^{\sigma }(E)$ is the transmission coefficient for spin channel $\sigma $. $T^{\sigma }(E)$ can be found from the Green's function between the leads\cite{Datta}. Calculations are done at zero magnetic field. In order to find spin separated solutions a small magnetic field, $\sim 0.05$T, is applied for the first $\sim $100 (out of 1000-20000) iterations. Although the direct effect of the magnetic field is very small it is sufficient to lift spin degeneracy and for the converged solution to be spin polarized. The temperature in all simulations was chosen to 1K.
%*************************************************
%******************* RESULTS *********************
%*************************************************
%

\section*{Results and Discussion}

\begin{figure}[tbh]
\resizebox{\linewidth}{!}{\includegraphics{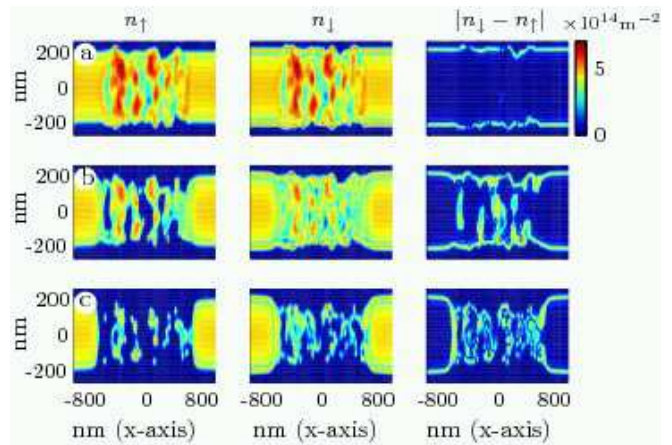}}
\caption{(color online) Electron densities $n^{\uparrow }$, $n^{\downarrow }$ (left and middle columns) and spin polarization $|n^{\downarrow }-n^{\uparrow }|$ (right column) for the 30nm-spacer sample at different gate voltages, $V_{g} =0$V, -0.03V, -0.07V (rows (a), (b), (c) respectively).} \label{fig:figure3} \end{figure}

The top panels in Fig.\ \ref{fig:figure3} %
%
%
%
%
%
%
%
%*************************************************
%**************** FIGURE THREE *******************
%*************************************************
 show the spin up/down electron densities and the spin polarization $|n^{\downarrow }-n^{\uparrow }|$ at $V_{g}=0$. For this gate voltage the impurities in the middle section cause a clear modulation of the electron density with a negligible spin polarization inside the wire. However, the electron density exhibits a pronounced spin polarization near the wire edges. Because incoming states in the leads are spin degenerate, the polarization along the edges was quite unexpected. To understand the origin of this spin polarization, we study an \textit{infinite} ideal homogeneous wire, where the confinement is modeled by a parabolic potential,
\begin{equation}
V_{par}(y)=V_{0}+\frac{m}{2}\left( \omega y\right) ^{2},
\label{eq:parabolicConfinement}
\end{equation}
$V_{0}$ being the bottom of the parabola. Note that the parabolic confinement represents an excellent approximation to the electrostatic potential from a gated structures\cite{Kumar, Reimann_RevModPhys, Ihnatsenka2005}. %*************************************************
%****************** FIGURE FOUR ******************
%*************************************************
\begin{figure}[tbh]
\resizebox{\linewidth}{!}{\includegraphics{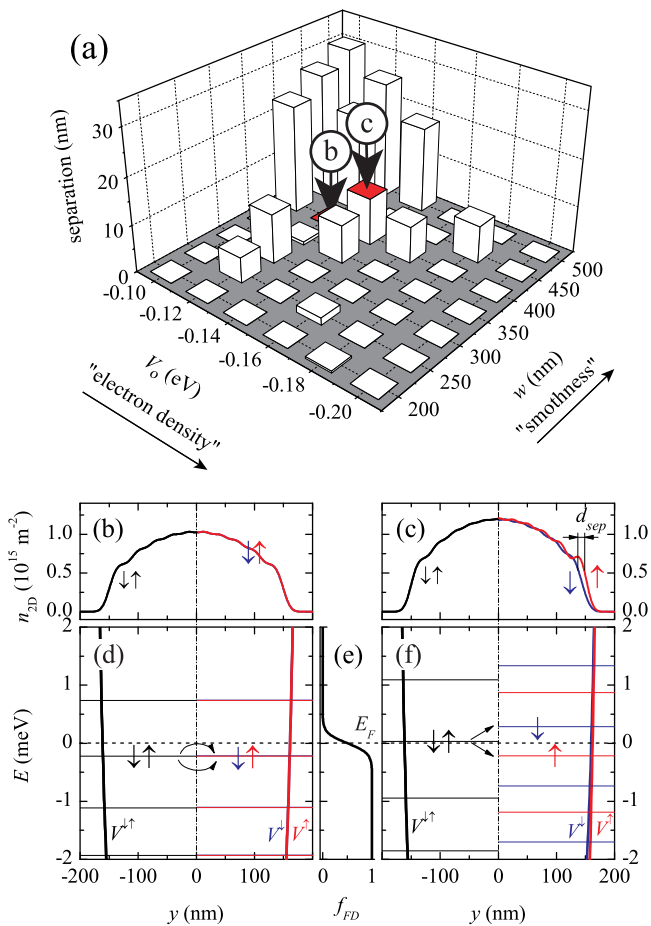}}
\caption{(color online) (a) Spatial spin separation, $d_{sep}$, at the boundary of the quantum wire vs the saddle point potential $V_{0}$ and the wire width $w$. $d_{sep}$ is loosely defined as the distance between spin species at the level 0.5$\times n(y=0)$. $V_{0}$ and $w$ can be ascribed as the electron density ant potential profile smoothness, respectively. In (b)-(f) the left and right panels show respectively the spin degenerate solutions (black lines) and spin-resolved solutions (red and blue lines). (b) and (d), electron density and band structure in the wire indicated \textcircled{b} in panel (a). (c) and (f), electron density and band structure in the wire indicated \textcircled{c} in panel (a).} \label{fig:figure4}
\end{figure}
At the same time, by changing the saddle point of the parabola, $V_{0}$, and the confinement strength, $\hbar \omega $, it is convenient to control both the electron density and the smoothness/steepness of the potential. Self-consistent solution of Eqs. (\ref{eq:hamiltonian})-(\ref{eq:electrondensity}) can be spin degenerate and spin polarized (left and right panels in Fig.\ \ref{fig:figure4} (b), (c)). As for the case of quantum wire of Fig.\ \ref{fig:figure1}, a small magnetic field was temporarily introduced to trigger spin degenerate solutions for some initial iterations. Figure \ref{fig:figure4}(c) shows a representative spin-polarized electron density in an ideal infinite quantum wire. As the electron density decreases at the edge of the wire, it becomes spin polarized and exhibits a spatial spin polarization yielding a separation, $d_{sep}$, between the spin up/down densities. This is summarized in Fig.\ \ref{fig:figure4} for a series of wire configurations. Along the $V_{0}$-axis in Fig.\ \ref{fig:figure4}(a) the width of the wire is held constant whereas the electron density grows as $|V_{0}|$ is increased (we keep the Fermi energy $E_{F}=0).$ Conversely, changing the width of the wire along the $w$-axis by decreasing the confinement strength, $\hbar \omega $, and keeping $V_{0}$ constant, a more shallow wire is studied. The behavior of the spatial spin polarization presented in Fig.\ \ref{fig:figure4}(a) shows that $d_{sep}$ increases as the electron density is decreased and the confinement becomes smoother. Note that this dependence of the $d_{sep}$ as a function of the electron density and the confinement strength is consistent with the corresponding behavior of $d_{sep}$ near the edges of a quantum wire in perpendicular magnetic field \cite{comp_strips}, where $d_{sep}$ also increases as the electron density is decreased and the confinement becomes smoother. It should be noted that in the present case of zero magnetic field, $d_{sep}$ shows a nonmonotonic dependence of the spin polarization and electron density/slope of confinement potential. This behavior of $d_{sep}$ is a manifestation of the subband structure in a quantum wire of a finite width. For example, if there is some energy level close to $E_{F},$ the exchange interaction effectively splits it into spin-up and spin-down subbands, see Fig.\ \ref{fig:figure4}(c),(f). As a result, the total charge density profile shows a spatial spin separation $d_{sep}$ at the boundary of quantum wire (left part of (c) in figure \ref{fig:figure4}). In contrast, a spin-degeneracy holds for energy levels far away from $E_{F},$ see Fig.\ \ref{fig:figure4}(b), (d).

Having established that an infinite quantum wire can have a spin-polarized solution, we conclude that this solution is triggered in the finite quantum wire as well, even though the electrons injected into the middle part of the wire are spin-degenerate (we stress that we always select the spin-unpolarized solution in the leads). A word of caution is however in order concerning a reliability of the above predictions for the spatial spin separation near the wire edges obtained within the DFT-LSDA. Our recent comparison of the DFT-LSDA and the Hartree-Fock (HF) approaches demonstrates that the two methods provide qualitatively (and in most cases quantitatively) similar results for electronic properties of ideal infinite quantum wires in the integer quantum\ Hall regime\cite{HFvsDFT}. However, in contrast to the HF approach, the DFT calculations predict much larger spatial spin separation near the wire edge for low magnetic fields (when the compressible strips for spinless electrons are not formed yet). Note that a comparative study of two methods can not distinguish which approach gives a correct result for $d_{sep}$ for zero field. This question can be resolved by a comparison to the exact results obtained by e.g. quantum Monte Carlo methods. We thus can not exclude that the predicted spin polarization near the wire boundaries as $B=0$ can be an artifact of the DFT-LSDA, and we defer this question to further studies.

Let us now focus on the spin polarization in the central part of the wire. Depending on the electron density, we can identify three regimes with qualitatively different behavior. In the first regime the spin polarization of the electron density $P_{n}=|n^{\uparrow }-n^{\uparrow }|/(n^{\uparrow }+n^{\uparrow })$ and the spin polarization of the conductance $  P_{G}=|G^{\uparrow }-G^{\uparrow }|/(G^{\uparrow }+G^{\uparrow })$ are negligible; for the 30nm spacer sample this is roughly between $-0.025$V$\lesssim V_{g}\lesssim 0$V (row (a) in Fig. \ref{fig:figure3}) while for the 60nm spacer sample this happens between $-0.045$V$\lesssim V_{g}\lesssim 0$V. In this regime the decreasing gate voltage causes a decreasing conductance (because of a reduction of the number of propagating subbands), but no significant spin polarization, except at the edges, occurs. The polarization at the edges is expected since the wire under consideration is wide ($\sim $500nm) and shallow (minimum potential $\sim -5$meV), which corresponds to the high polarization region of Fig. \ref{fig:figure4}(a). The self-consistent potential, shown for a slice along the middle of the wire in Fig. \ref{fig:figure2}(a), (b) is well below the Fermi energy, such that the characteristic potential fluctuations of the long-range impurity potential are much smaller than the average distance from the potential bottom to $E_{F}$. %
%*************************************************
%***************** FIGURE FIVE *******************
%*************************************************
\begin{figure}[tbh]
\resizebox{7cm}{!}{
\includegraphics{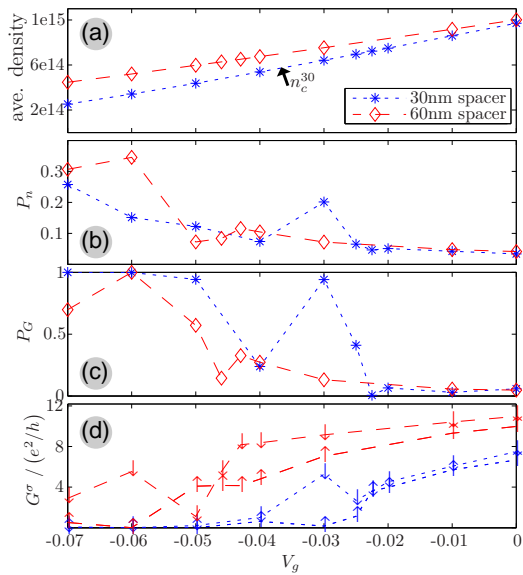}}
\caption{(color online) (a) The average electron density, $\bar{n}^{\uparrow }+\bar{n}^{\downarrow }$, directly beneath the top gate. The arrow indicate the critical electron density, $n_{c}$, for the MIT according to Eq.\ (\protect\ref{eq:Efros1993}) for the 30nm spacer sample. For the 60nm spacer sample $n_{c}=3.0\times 10^{14}$m$^{-2}$ is achieved for a gate voltage lower than -0.07 V. (b) The density spin polarization, $P_n$; (c) the spin polarization of the conductance, $P_G$; (d) the spin-resolved conductance $G^{ \protect\sigma }$.} \label{fig:figure5}
\end{figure}

As the gate voltage becomes more negative, the wire undergoes a spin polarization in the central part, see Fig.\ \ref{fig:figure5}(b), (c). For the 30nm spacer sample (row (b) in figure \ref{fig:figure3}) this occurs for the gate voltages $V_{g}\lesssim -0.025$V and for the 60nm sample, $V_{g}\lesssim -0.045$V.
%In %this voltage span the conductance for one of the spin species, $G^{\sigma }$% %, drops to zero (Fig.\ \ref{fig:figure5}(d)), indicating that the confining %potential is close to the Fermi energy in many regions of the wire and that %pinch off is imminent. Spin polarization in this regime is local and occurs %on multiple places throughout the impurity section, Fig.\ \ref{fig:figure3}% %(b) rightmost panel. Fig.\ \ref{fig:figure2}(c) and (d) shows that the %confining potential for one of the spin species (sometimes spin-up, %sometimes spin-down) is completely above the Fermi energy in sections of the %wire meaning that there is 100\% spin polarization in these regions. [% %\textbf{this is not quite clear to me: if you have region above }$E_{F},$% %\textbf{you would expect that the corresponding density = 0, i.e. no spin %polarization}]. We interpret this as an effect of the exchange energy being %sufficiently large to completely suppress one spin species in a region of %the wire if the confining potential is close to the Fermi energy. The %spin-up/down densities remain almost degenerate until the exchange energy $% %|E_{xc}|>E_{f}-V^{\sigma }$.
%
The splitting results in the fragmentation of spin-up/down densities into spin-polarized islands in the wire seen in row (b) of figure \ref{fig:figure3}. The onset of spin polarization is displaced towards a lower gate voltage for the thicker spacer, Fig.\ \ref{fig:figure5}(b/c), but it is not clear whether there is any other qualitative difference between the two samples. To settle this would require further computations over more samples with varying spacer thickness and donor sheet configurations. However, because of the extensive computational efforts needed to find a convergent solution (sometimes requiring up to 20000 iterations) we were in position to study only two representative wires with spacers 30 nm and 60 nm.

Finally we identify a third, non-conducting regime, corresponding to a metal-insulator transition (MIT); $V_{g}\lesssim -0.04$ V for the 30nm spacer and $V_{g}\lesssim -0.07$ V for the 60nm spacer. Conductance is pinched off and electrons are trapped in isolated pockets along the wire, Fig.\ \ref{fig:figure3} (c). This electron-droplet state has been analyzed thoroughly in \cite{Tripathi, Efros1993, Shi, Fogler}. Using arguments based on the screening of the impurity potential by the 2DEG, Efros \emph{et al.\ }\cite{Efros1993} gave an expression for the critical density, $n_{c}$, where the metal-insulator transition (MIT) occurs,
\begin{equation}
n_{c}=\beta \frac{\sqrt{\rho _{d}}}{d_{spacer}},  \label{eq:Efros1993}
\end{equation}%
where $\beta $ is a numerical constant =0.11 and $\rho _{d}$ the donor density. We use this expression to find the approximate point for the MIT in our samples. For the 30nm spacer Eq.\ (\ref{eq:Efros1993}) yields $ n_{c}^{{}}=5.9\times 10^{14}$m$^{-2}$ (see Fig.\ \ref{fig:figure5} (a)) and for the 60nm spacer $n_{c}=3.0\times 10^{14}$m$^{-2}$. The predictions for $n_{c}$ agree rather well with the numerical results that show that the wire undergoes a transition to a spin-polarized regime before the MIT occurs (see Fig.\ \ref{fig:figure5}) (b), (c). Thus, close to the pinch-off regime, the DFT-LSDA predicts formation of the spin-polarized electron lakes trapped in the minima of the long-range impurity potential. Such localized states might be relevant to the experimental observations of Bird \textit{at al.} that provide evidence of bound state-mediated resonance interaction between the coupled quantum point contacts close to the pinch-off regime\cite{Bird_QPC}. While microscopic origin on the effect is still under debate, our findings indicate that because the spin-polarized localized states trapped in minima of the impurity potential are generic feature of modulation-doped split-gate heterostructures, they might be relevant for the interpretation of the effect reported by Bird \textit{et al.} \cite{Bird_QPC}.

Finally, a comment is in order concerning applicability of the method used. In our calculations we assume a constant chemical potential throughout the system. This condition is certainly violated in the regime close to the pinch off when the electrons are trapped in isolated lakes containing an integer number of charges (i.e. in the Coulomb blockade regime of electron transport). It is now well recognized that the standard DFT-LSDA is not expected to work in the Coulomb blockade regime of weak coupling because of the spurious self-interaction errors caused by the lack of the derivative discontinuity of the exchange and correlation potentials in the standard DFT\cite{Toher}. The validity of the present method is limited to the case of strong coupling when electron number in the structure is not quantized (i.e. the Coulomb charging is unimportant) and the conductance of the systems exceeds the conductance unit $G_{0}=2e^{2}/h$ (see Ref. \cite{Ihnatsenka2007} for a detailed discussion and further references). Because of the uncorrected self-interaction of the standard DFT-LSDA, the electron lakes in the pinch-off regime do not contain an integer electron number and the calculated conductance does not exhibit expected Coulomb-blockade peaks. We thus conclude that while the present calculations qualitatively capture the onset of the pinch-off regime, a quantitative description of this regime (accounting for the quantized electron number in the pockets as well reproducing the Coulomb blockade peaks in the conductance) would require methods that go beyond the standard DFT-LSDA scheme utilized in the present calculations.

Let us now compare our findings with available experimental results. Spontaneous spin-polarization at low electron densities has been probed in various systems, Refs. \cite{Ghosh2004, Ghosh2007, Goni2007, Prus, Shashkin}. In \cite{Ghosh2004}, Ghosh \emph{et al.} studied the evolution of the zero bias anomaly (ZBA)\cite{Cronenwett2002} in 2DEGs for low and zero magnetic fields. The behavior of the ZBA was associated with different spin states in the 2DEG and measurements over different disorder configurations (cool downs) and temperatures indicated that the spin polarization observed is a generic effect for low density 2DEGs. The ZBA was most easily observed in a small disorder window between the metallic and insulating regime. This is qualitatively consistent with the window of high spin polarization we find above (Fig.\ \ref{fig:figure5}(b)-(d)). Further experiments on the ZBA in 2DEGs\cite{Ghosh2007} suggested the formation of localized magnetic moments due to spin polarized regions in the 2DEG as the electron density is lowered. This was understood as an effect of the potential due to background disorder which is similar to the impurity induced spin polarized droplets we find in Fig.\ \ref{fig:figure3} (third column). Many of the observations in \cite{Ghosh2007} were strongest for electron densities around $1-3\times 10^{14}$m$^{-2}$, a slightly lower electron density than we find. In our case the \emph{average} density below the top gate for the onset of the spin polarization is 6.4$\times 10^{14}$m$^{-2}$ for the 30 nm spacer case and $5.1\times 10^{14}$m$^{-2}$ for the 60 nm spacer case.

A direct measure of magnetization of the 2DEG at low electron densities were done in\cite{Prus, Shashkin} for Si-SiO$_2$ heterostructure 2DEGs. By modulating an in plane magnetic field and measuring the minute current between gate and 2DEG the thermodynamic magnetization of the 2DEG is found through Maxwell's equations\cite{Prus, Shashkin}. Both Prus \emph{et al.}\cite{Prus} and Shaskin \emph{et al.}\cite{Shashkin} find that the spin susceptibility is critically enhanced prior to the metal-insulator transition in the 2DEG. It is, however, not clear from the experiment whether a spin polarized phase actually exists between the metal and insulating phase or if there is only an increased magnetization in the metallic phase.

Resonant inelastic light scattering measurements on GaAs single wells showed direct evidence for spin polarization at low densities\cite{Goni2007}. Calculations using time-dependent local spin-density approximation in the same paper predicted a stable polarized state below an electron density of $3.4\times 10^{14}$m$^{-2}$. This is once again slightly lower than we encounter.

\subsection*{Conclusions}

Using the spin density functional theory we have studied spin-polarization of a 2DEG in split-gate quantum wires formed in modulation-doped GaAs heterostructures focusing on the effect of the long-range impurity potential originating from the remote donors. We find that depending on the electron density, the spin polarization exhibits qualitatively different features in three different regimes. For the case of relatively high electron density, when the Fermi energy $E_{F}$ exceeds a characteristic strength of a long-range impurity potential $V_{donors}$, the density spin polarization inside the wire is practically negligible and the wire conductance is spin-degenerate. We find however a strong spin polarization near the wire boundaries. When the density is decreased such that $E_{F}$ approaches $V_{donors},$ the electron density and conductance quickly become spin polarized. With further decrease of the density the electrons are trapped inside the lakes (droplets) formed by the impurity potential and the wire conductance approaches the pinch-off regime\textbf{.} Experimentally, spin polarization prior to localization of the 2DEG has been suggested in\cite{Ghosh2004, Goni2007, Ghosh2007}. The electron density where we find spin polarization in the wire is roughly equal to what has been determined experimentally in GaAs/AlGaAs\cite{Goni2007, Ghosh2007}. Direct measurements of the magnetization of the 2DEG in Si-SiO$_2$ heterostructures suggests an increased spin susceptibility\cite{Prus, Shashkin} close to the MIT but it is not clear in these experiments whether an spin polarized phase, as we find, exists.
%*************************************************
%***************** BIBLIOGRAPHY ******************
%*************************************************

\end{document}